\PassOptionsToPackage{fleqn}{amsmath}
\documentclass[twocolumn,preprintnumbers,amsmath,amssymb]{revtex4}
\usepackage{graphicx}
\usepackage{dcolumn}
\usepackage{bm}
\usepackage{amssymb}
\usepackage{amsmath}

\begin{document}
\title{Electrochemical Oxygen Intercalation into Sr$_2$IrO$_4$}

\author{L. Fruchter$^{1}$, V. Brouet$^{1}$, D. Colson$^{2}$, J.-B. Moussy$^{2}$, A. Forget$^{2}$, Z.Z. Li$^{1}$}%
\affiliation{$^1$Laboratoire de Physique des Solides, C.N.R.S., Universit\'{e} Paris-Sud, 91405 Orsay, France}
\affiliation{$^2$Service de Physique de l'Etat Condens\'{e}, CEA-Saclay, 91191 Gif-sur-Yvette, France}

\date{Received: date / Revised version: date}

\begin{abstract}{Oxygen was electrochemically intercalated into Sr$_2$IrO$_4$ sintered samples, single crystals and a thin film. We estimate the diffusion length to a few $\mu$m and the concentration of the intercalated oxygen to $\delta$ $\simeq$ 0.01. The latter is thus much smaller than for the cuprate and nickelate parent compounds, for which $\delta$ $>$ 0.1 is obtained, which could be a consequence of larger steric effects. The influence of the oxygen doping state on resistivity is small, indicating also a poor charge transfer to the conduction band. It is shown that electrochemical intercalation of oxygen may also contribute to doping, when gating thin films with ionic liquid in the presence of water.}
\end{abstract}
%

%
\maketitle

The spin-orbit-induced Mott insulator Sr$_2$IrO$_4$ has been the subject of numerous studies in the recent years. For this compound, although extended 5d orbitals reduce the electron-electron interaction, as compared to the 3d transition metal compounds like cuprates, strong spin orbit coupling associated to the heavy Ir is thought to restore the impact of electronic correlations. This yields a spin-orbital Mott insulator, based on a J$_{eff}$ = 1/2 state obtained from the combination of 5d orbitals and spin \cite{Kim08,Jackeli09}. The similarities with the cuprate La$_2$CuO$_4$ with a similar crystallographic structure\cite{Crawford94} (except for a large rotation of the IrO$_6$ octahedra) was early pointed out. As for La$_2$CuO$_4$, the insulating Sr$_2$IrO$_4$ is antiferromagnetically ordered in the IrO$_2$ plane, with a ferromagnetic moment associated to the octahedra rotation\cite{Vaknin87,Kim09,Ye13}. It is believed that doping should yield a particle-hole phase diagram similar to the one of the cuprates \cite{Wang11}, including the possibility of a superconducting phase\cite{Yan15,Zhao16,Kim16}. Thus, the stakes at doping the IrO$_2$ plane are high, and this was reported being done in several ways: chemical substitution \cite{Klein08,Ge11,Yan15,Brouet15}, electric field effect\cite{Ravichandran13,Lu15}, and oxygen stoichiometry\cite{Korneta10}. However, the doping level obtained with these techniques is low: the field effect fails to reach a metallic state, while chemical substitution barely reach it, and is limited to 4-5\%.

Oxygen doping has for long been an attractive method, as relatively high mobility of the oxygen atom in the perovskite materials allows in principle for reversible adjustment of stoichiometry. The method, however, had limitations in the case of La$_2$CuO$_4$, requiring high oxygen pressure annealing to obtain superconducting samples with $\delta$ $\simeq$ 0.03 (with a phase separation below room temperature into two phases $\delta$ $\simeq$ 0.01 and $\delta$ $\simeq$ 0.06 - for a review, see Ref.~\onlinecite{Chou96} and refs. therein). Chemical substitution was thus usually preferred to scan the particle-hole phase diagram over a large doping range\cite{Johnston91}. Concerning Sr$_2$IrO$_{4+\delta}$, besides the work in Ref.~\onlinecite{Korneta10}, which reports dramatic doping effects for stoichiometry variation $\delta \approx$ -0.04, we are not aware of any other similar report. We attempted to alter the oxygen contents of Sr$_2$IrO$_4$ powders, by firing them under vacuum for several days (650 $^\circ$C), or under N$_2$ or O$_2$ atmosphere (600-800 $^\circ$C). No weight variation of a sintered sample of 2.4 g in a thermobalance with a working sensibility of 0.01 mg could be detected, indicating $\delta$ $<$ 10$^{-4}$.

\begin{figure}
\resizebox{0.7\columnwidth}{!}{%
  \includegraphics{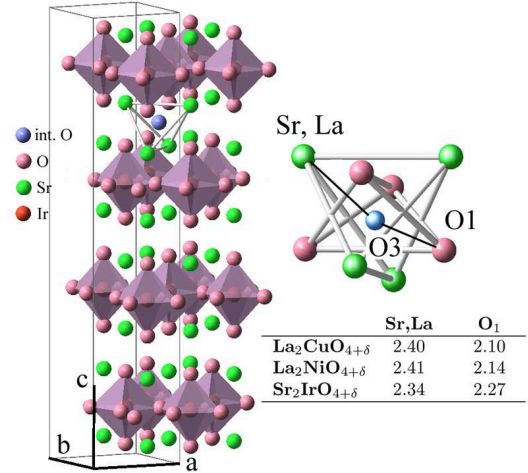}
  }
\caption{Representation of the interstitial oxygen atom in Sr$_2$IrO$_{4+\delta}$, assuming it is similar to the one in La$_2$CuO$_{4+\delta}$ and La$_2$NiO$_{4+\delta}$. The atomic distances in table are the ones for the undistorted structure to O$_3$ ($\frac{1}{4}$0$\frac{1}{4}$) (averaged, in the case of La$_2$CuO$_4$ and La$_2$NiO$_4$, for which there is a tilt of the oxygen octahedra).}\label{structure}
\end{figure}

The higher efficiency of electrochemical methods to insert oxygen into oxides is known for long. For La$_2$CuO$_{4+\delta}$, homogeneous hole doping as large as $\delta$ = 0.12 was obtained in this way, yielding the metallic, superconducting phase\cite{Grenier91,Grenier92,Radaelli93}. A defect structure model for La$_2$NiO$_{4+\delta}$ was found also appropriate for La$_2$CuO$_{4+\delta}$, where the interstitial oxygen resides between the two LaO layers, and is coordinated to four near-neighbor La atoms, inducing a displacement of four neighbor apical oxygen\cite{Jorgensen89,Chaillout90}. LaO bi-layers in La$_2$NiO$_4$ and La$_2$CuO$_4$ should be essentially similar to the SrO layers in Sr$_2$IrO$_4$, with the reservation that the larger Sr$^{2+}$ might make the insertion of the interstitial oxygen more difficult than for La$^{3+}$. Indeed, taking into account the larger ionic radius for Sr$^{2+}$ than for La$^{3+}$, the distortion of the Sr tetrahedron in Sr$_2$IrO$_4$, needed to accommodate an interstitial oxygen, as in Fig.~\ref{structure}, may be estimated 3 times larger than for the La one, in La$_2$CuO$_4$.

To study the electrochemical insertion of oxygen in Sr$_2$IrO$_4$ samples, we have used sintered powders, single crystals, and a thin film. The 88 nm-thick film was epitaxially grown on (001) SrTiO$_3$, by pulsed laser deposition, at 600 $^\circ$C and oxygen pressure 5 10$^{-2}$ mbar. The single crystals were cleaved from large crystals grown using a self-flux technique in platinum crucibles, similar to the one in Ref.~\onlinecite{Kim09}. Typical dimensions of such crystals was 1000 x 300 x 50 $\mu m^3$. The cleaved (001) surface of the samples was optically flat, sometimes with a few steps. According to Ref.~\onlinecite{Dai14}, the crystals cleave between SrO layers, leaving a neutral surface. These samples were appropriate to obtain voltamogramms and for kinetics studies, which were found reproducible for several samples. The powders were elaborated from stoichiometric IrO$_2$ (99.99 \% in purity) and dried SrCO$_3$ (99.99\% in purity) that were mixed and pre-reacted at 700 $^\circ$C in an alumina crucible for 24 hours to decompose the carbonate. The resulting materials were reground, pelletized, and fired for 48 h. at 1100 $^\circ$C, twice. The ground powder was finally pressed into pellets and sintered at 1150 $^\circ$C, yielding a typical grain size $\lesssim$ 1 $\mu m$. Samples were contacted by simply pressing a 25 $\mu m$ diam. gold wire to their surface. Although we found it is also possible to contact single crystals using silver paste, provided the contacts are buried into epoxy resin to prevent contact from the electrolyte, the use of mechanical contacts was preferable to strictly rule out any contribution from the contacts oxidation. The powder samples were not suitable to measure resistivity before and after electrochemical treatment, due to their loose mechanical cohesion and their tendency to crumble after treatment, but we could gain this information from the thin film. The electrochemical measurements and cycling were done in a conventional three-electrode cell. The auxiliary electrode was a gold one, and the reference electrode was Hg/HgO filled with KOH at the same concentration as the cell electrolyte, or Hg/HgSO$_4$ for H$_2$SO$_4$ electrolyte. The working electrodes were not rotated. All potentials are referred to SHE (standard hydrogen electrode).

\begin{figure}
\resizebox{0.95\columnwidth}{!}{%
  \includegraphics{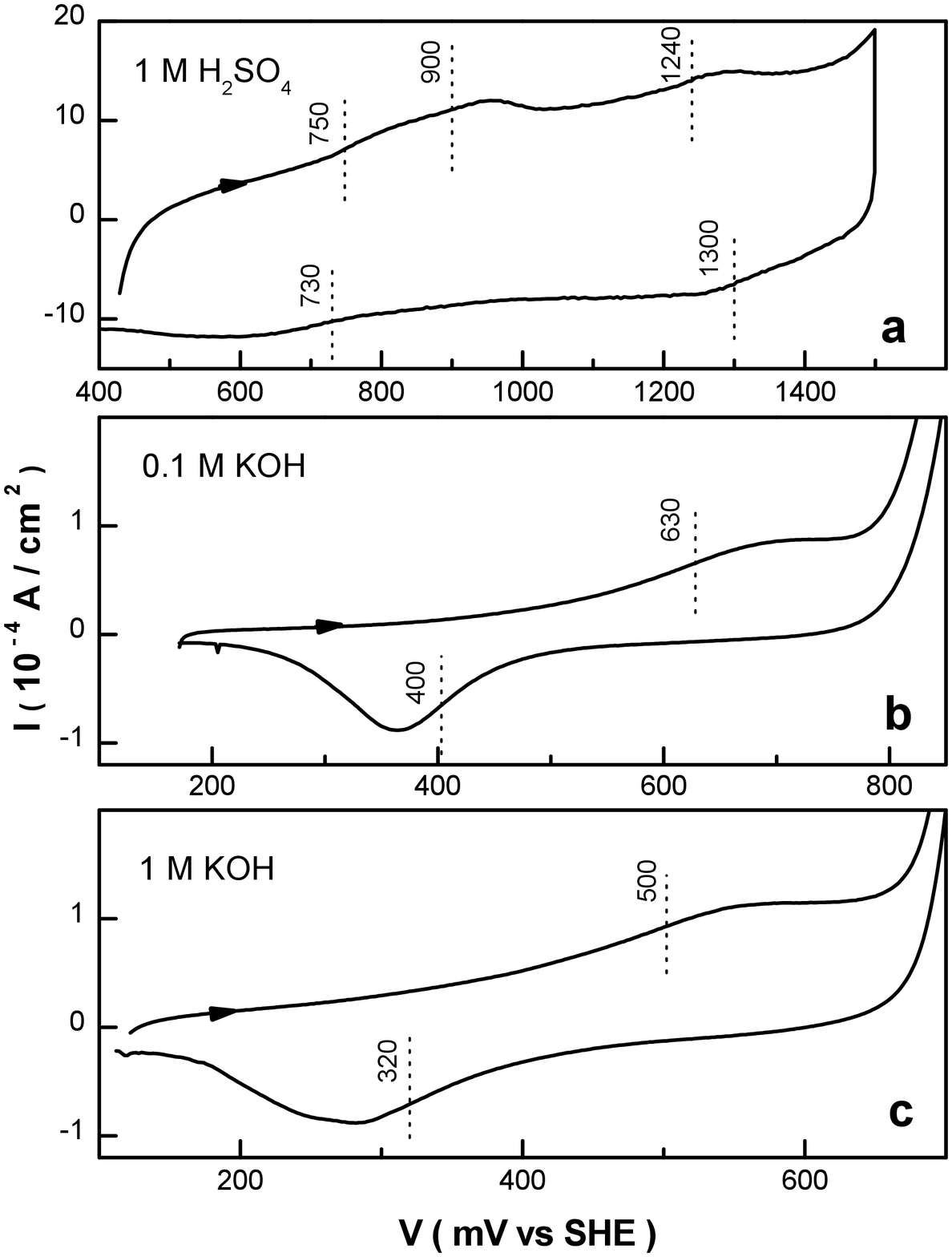}
  }
\caption{Cyclic voltammograms for a cleaved (001) Sr$_2$IrO$_4$ single crystal surfaces (from top to bottom: $dV/dt$ = 50, 10, 14 mV/s, pH = 0.3, 13, 14). The dotted lines indicate the steepest slope location of the redox peaks.}\label{cycles}
\end{figure}

Fig.~\ref{cycles} displays typical voltamogramms obtained in acid (H$_2$SO$_4$) and alkaline (KOH) electrolyte. Repeated electrochemical cycles as in Fig.~\ref{cycles}c (which are the conditions used for oxygen intercalation in the following), left the crystal surface shiny, and there was no trace of corrugation that could be detected by inspection with scanning electron microscopy. The composition of the surface was also found unchanged within energy-dispersive X-ray analysis accuracy (variations less than 1 \%), and there was no detectable K/Sr substitution (K/Sr $<$ 6 10$^{-4}$ within typical probe depth 1 $\mu$m from the crystal surface of anodized samples). The voltamograms are similar to the ones obtained for 3d transition-metal oxides\cite{Grenier91,Grenier92}. In 1 M KOH, following Refs.~\onlinecite{Goodenough90,Grenier92}, we find: an oxidation peak at 500 mV, corresponding to the overall reaction:
\begin{equation}
\textrm{Sr}_2\textrm{IrO}_y + 2\delta \, \textrm{OH}^- \longrightarrow \textrm{Sr}_2\textrm{IrO}_{y+\delta} + \delta \, \textrm{H}_2\textrm{O} + 2\delta \, \textrm{e}^-
\label{oxidation}
\end{equation}

oxygen evolution above 700 mV:
\begin{equation}
\textrm{OH}^- \longrightarrow 1/2 \, \textrm{H}_2\textrm{O} + 1/4 \, \textrm{O}^{\nearrow}_{2} + \, \textrm{e}^-
\label{evolution}
\end{equation}

and a reduction peak at 320 mV. When driven far above the oxygen evolution potential, the system may display an additional peak on the cathodic sweep, which we interpret as the contribution of O$_2$ present on the sample surface. An oxydation peak, rather than a plateau as in Fig.~\ref{cycles}, could only be observed for more reversible cycles on rough sample surfaces, at low sweep speed, when the oxidation peak and the oxygen evolution exhibited a larger potential difference (see discussion below).

The oxidation potential is smaller by approx. 100 mV than for La$_2$CuO$_4$ in the same conditions. The difference between the oxidation and reduction potentials is large ($\approx$ 180 mV). This large irreversibility and the shallowness of the oxidation peak make the determination of a thermodynamic standard potential difficult, which we however estimate from the average of the oxidation and reduction potentials as 410 mV.
As shown in Ref.~\onlinecite{Grenier92}, the competition between reactions \ref{oxidation} and \ref{evolution} is actually a competition between the oxidation of the bulk metal (with diffusion of the interstitial oxygen):

\begin{equation}
2\, \textrm{Ir}^{4+}_{bulk} +\, \textrm{Ir}^{4+}_{surf}-\textrm{OOH}^{-} \longrightarrow 2\, \textrm{Ir}^{5+}_{bulk} +\, \textrm{O}^{2-}_{i} +\, \textrm{Ir}^{4+}_{surf}-\textrm{OH}^{-}
\label{evolution1}
\end{equation}

and of the surface one:

\begin{equation}
\textrm{Ir}^{4+}_{surf}-\textrm{OOH}^- +\, \textrm{OH}^- \longrightarrow \textrm{Ir}^{5+}_{surf}-(\textrm{O}_2)^{2-} +\, \textrm{H}_2\textrm{O} +\, e^-
\label{evolution2}
\end{equation}

While the manifestation of the Ir$^{5+/4+}$ couple \textit{at the surface} of the samples in acidic and alkaline media has been observed for long (see e.g. Ref.~\onlinecite{Goodenough90}), Eq.~\ref{evolution1} shows that the kinetics of the interstitial oxygen diffusion is essential to determine whether the oxidation of the metal also occurs in the bulk, with the intercalation of the oxide ions. It is clear that we do not have a direct evidence for the bulk oxidation state change (which would be challenging, as we show that only a minute amount of oxygen is introduced). However, the observation below of a bulk doping effect on thin films, and the evidence for a modification of the material structural parameters, are strong indications that oxidation in the bulk as in Eq.~\ref{evolution1} occurs.

\begin{figure}
\resizebox{0.95\columnwidth}{!}{%
  \includegraphics{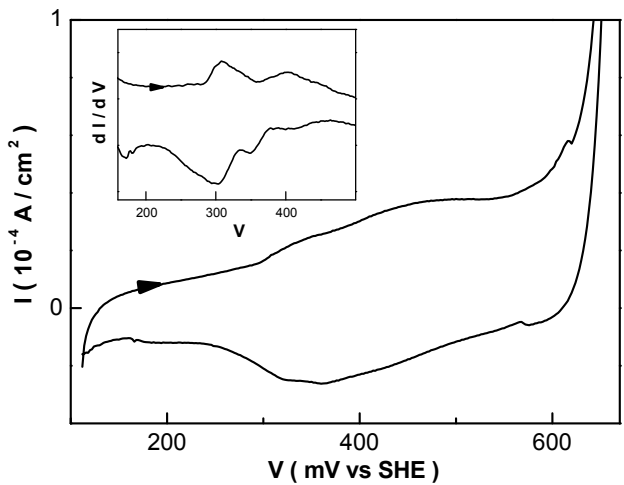}
  }
\caption{Splitting of the Ir$^{5+/4+}$ couple peaks observed at low sweep speed ($5 \, 10^{-4}$ V/s) on a cleaved (001) Sr$_2$IrO$_4$ single crystal surface (1 M KOH).}\label{doublepeak}
\end{figure}

\begin{figure}
\resizebox{0.95\columnwidth}{!}{%
  \includegraphics{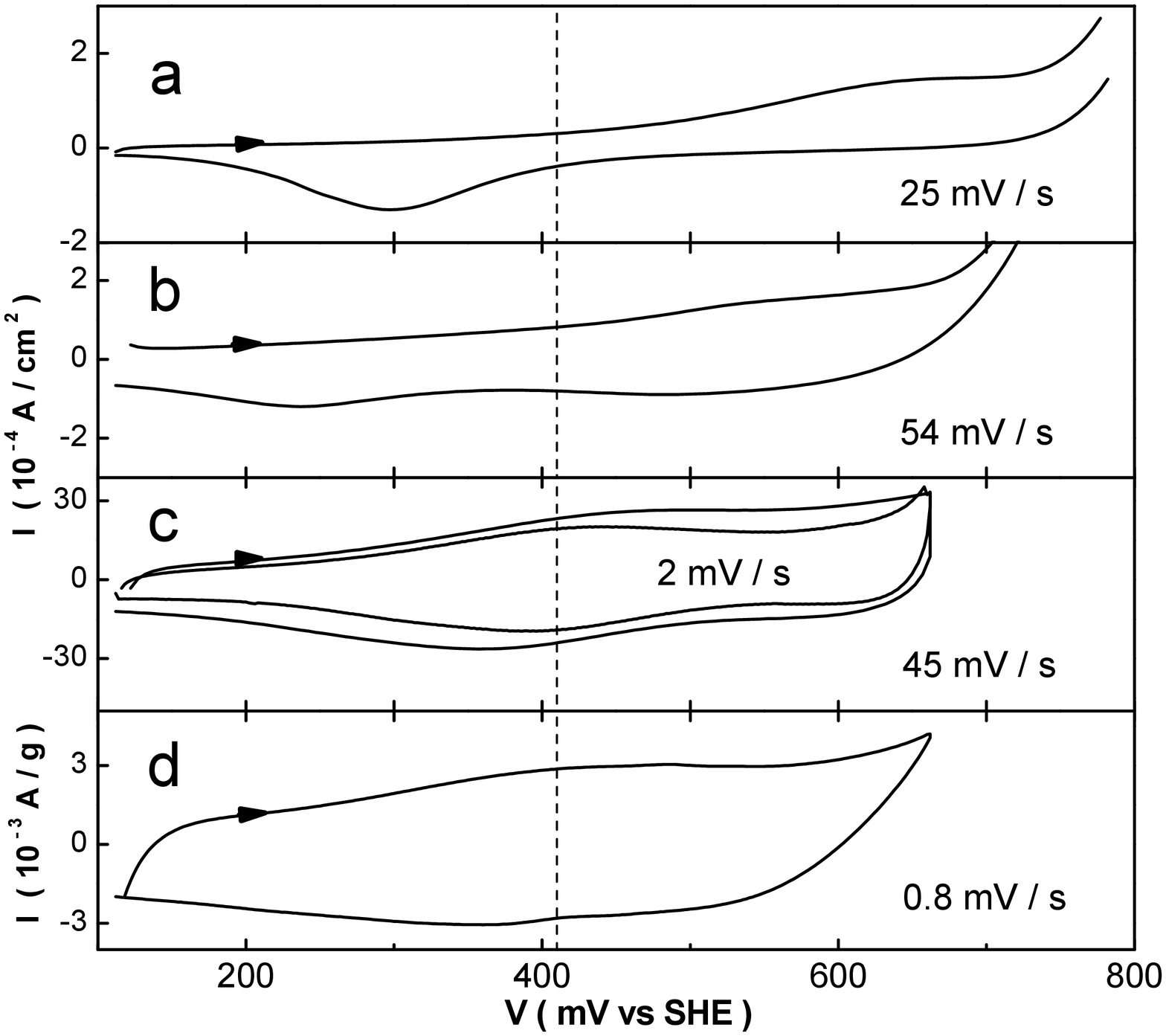}
  }
\caption{Cyclic voltammograms for different Sr$_2$IrO$_4$ samples (1 M KOH). a: cleaved single crystal, b: thin film, c: abraded single crystal, d: sintered powder. The dotted line at 410 mV is the estimated Ir$^{5+/4+}$ redox potential.}\label{comparison}
\end{figure}

From Eq.~\ref{evolution2}, the Ir$^{5+/4+}$ couple is expected to shift with pH. The couple potential at $\approx$ 1270 mV for 1 M H$_2$SO$_4$ being shifted from the one at $\approx$ 410 mV for 1 M KOH by $\approx$ -60 mV/pH = $-2.3 RT/F$ (Fig.~\ref{cycles}), as expected for a quasi-reversible Nernstian reaction as Eq.~\ref{evolution2}, these two couples are likely the same. Minguzzi et al computed the redox potential for Ir$^{5+/4+}$ in 1 M H$_2$SO$_4$ to be $\approx$ 1240 mV\cite{Minguzzi14}, so we assign this couple to Ir$^{5+/4+}$. Then, the redox couple at $\approx$ 800 mV for 1 M H$_2$SO$_4$ in Fig.~\ref{cycles} is likely Ir$^{4+/3+}$. In the bulk, this couple would imply oxygen \textit{vacancies}. We notice that there is a splitting of both the anodic and the cathodic peaks in 1M H$_2$SO$_4$. A similar splitting may be observed also for the Ir$^{5+/4+}$ couple in 1 M KOH (Fig.~\ref{doublepeak}), the origin of which is not known. The irreversibility of the redox reaction was found strongly dependent upon the surface state of the sample. In 1 M  KOH, atomically flat surfaces, as likely the case for cleaved single crystals and thin films, exhibited irreversibility as large as $\approx$ 200 mV (Fig.~\ref{comparison}a,b), while abraded single crystal surface and sintered powder showed much smaller irreversibility. The redox peaks were in this case obscured by the large double layer capacitance for these rough surfaces (Fig.~\ref{comparison}c,d). The larger irreversibility for cleaved crystals and films is likely due to larger activation energy, yielding larger overpotential for the redox reactions. This could be due to the presence of the SrO termination layer acting as a barrier for the redox reactions, while abraded surfaces directly expose reactive Ir sites, and so allow faster and more reversible kinetics.

\begin{figure}
\resizebox{0.95\columnwidth}{!}{%
  \includegraphics{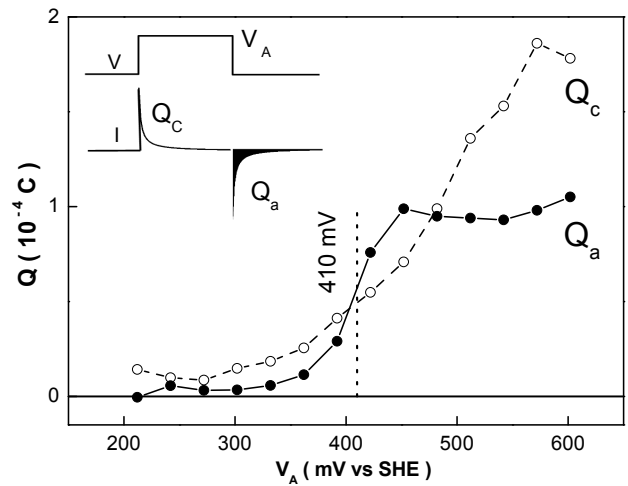}
  }
\caption{Cathodic charge (full symbols), and anodic charge (empty symbols), as a function of the anodic charging potential, for a cleaved (001) Sr$_2$IrO$_4$ single crystal ($7.6 \, 10^{-3}$ cm$^2$ x $50 \,\mu$m) in 1 M KOH.}\label{charge}
\end{figure}

We estimated the amount of oxygen incorporated by the electrochemical reaction using the following coulometric procedure. The samples were first charged at a potential above the oxidation one, and allowed to react until the working electrode current decayed to negligible values. Then, the potential was set to a value well below the reduction peak, and the current was monitored to obtained the total charge associated to the chemical species intercalated into the sample (Fig.~\ref{charge}, inset). In this way, we make sure that the competing oxygen evolution reaction (Eqs.~\ref{evolution}) does not contribute to the estimated charge, as the products of the reaction for the inverse reduction reaction are not available at the electrode. Fig.~\ref{charge} displays the charge obtained in this way, as a function of the oxidation potential ($V_A$). It is seen that the charge steeply increases when the sample is anodized above 410 mV, which is our previous estimation for the thermodynamic potential. Conversely, no saturation is observed for the anodic charging, as a consequence of the competing oxygen evolution reaction. Then, the saturation charge value for crystals is found proportional to the surface area of the samples, 1.2 10$^{-2}$ C cm$^{-2}$ 
(this is found reproducible for cleaved surfaces - see supplement S1, showing that it is indeed a surface charge, and intrinsic for the (001) surface of the crystals). The saturation charge for sintered samples is found proportional to the mass of the sample, as would be expected for a homogeneous intercalation independent of the grain size, 2.1 10$^3$ C mol$^{-1}$. In the case of the crystals, the surface charge cannot be strictly restricted to the cristallographic surface. Indeed, assuming a full occupation of the interstitial sites (as depicted in Fig.~\ref{structure}), the interstitial density would be 2.2 10$^{-9}$ mol cm$^{-2}$. Assuming O$^{2-}$ oxidation state, this yields the surfacic charge 4.2 10$^{-4}$ C cm$^{-2}$. The homogeneity of our results for different samples rules out that some extrinsic surface roughness could account for this discrepancy, and oxygen must be assumed to diffuse into the crystal. (A similar conclusion cannot be drawn for the Ir$^{4+/3+}$ couple in acid solution, as we evaluate the charge involved in this case to only 2 10$^{-4}$ C cm$^{-2}$, thus pointing to the reduction of the surface Ir, rather than to the creation of vacancies in the bulk). Assuming that the polycrystalline samples are homogeneously intercalated and representative of the single crystal diffusion region, the anodized samples composition is found Sr$_2$IrO$_{4+\delta}$, with $\delta$ = 0.01 (with O$^{2-}$ oxidation state assumption. In the rest of the text, $\delta$ is the oxygen interstitial content \textit{variation} involved in the electrochemical cycles, associated to the Ir$^{5+/4+}$ couple, and we do not consider a possible stoichiometry offset, which should strictly be written as Sr$_2$IrO$_{4+y+\delta}$). With this concentration of interstitial oxygen, assuming a simple linear concentration diffusion profile at the crystal surface, we obtain the diffusion length $l_D \simeq 2 \, a^{*3} C / (2e) \delta$ $\simeq$ 5 $\mu m$ ($a^*$ = 3.95 $\AA$ is the pseudo cubic unit cell, $C$ is the surfacic charge, and $2e$ is the assumed $O$ oxydation state). This validates \textit{a posteriori} the homogeneous hypothesis for polycrystalline samples (see S2, for a discussion on the uncertainty of this figure). The interstitial oxygen concentration obtained here is well below that reported for La$_2$CuO$_{4+\delta}$ ($\delta$ = 0.07 and 0.12 in Refs.~\onlinecite{Grenier91},~\onlinecite{Radaelli93} respectively).

\begin{figure}
\resizebox{0.95\columnwidth}{!}{%
  \includegraphics{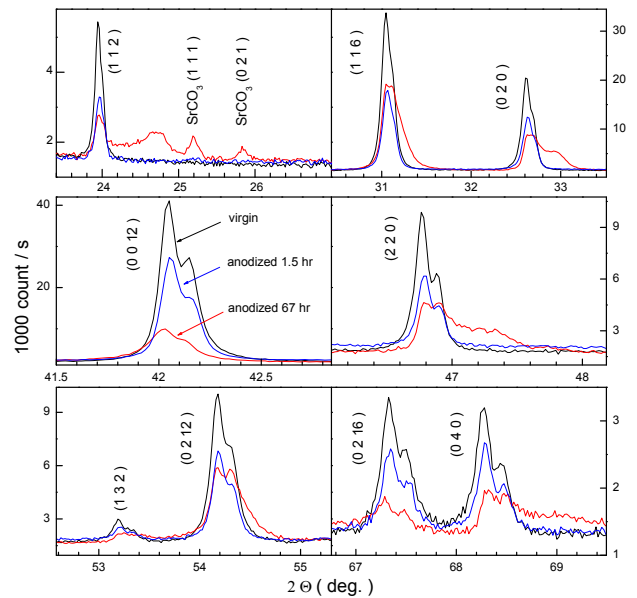}
  }
\caption{X-ray diffraction spectra, prior and after anodizing a sintered Sr$_2$IrO$_4$ sample at 510 mV in 1 M KOH for 1.5 hr (blue curve) and 67 hrs (red curve).}\label{XRDsintered}
\end{figure}

Attempts to further increase the oxygen contents of the sintered samples, using extended anodizing time, yielded a decomposition of the material. As an example, samples treated over times as long as 67 hrs displayed a marked degradation of the cristallinity, with additional diffraction peaks and the presence of SrCO$_3$, as evidenced by powder diffraction scans (Fig.~\ref{XRDsintered}), which likely originates from the carbonation with dissolved or atmospheric CO$_2$ of the damaged compound. On the other hand, at short times (typ. 1 hr), once the anodic current has dropped to its background steady value and the charging plateau as in Fig.~\ref{charge} is reached, there is no sign of sample degradation from the X-ray analysis.
The stability of the compound in alkaline conditions is further confirmed by electronic microscopy analysis (supplement S3). While sintered samples soaked a few hours in acidic conditions display etched grains, the ones soaked in alkaline conditions show no morphological change. Also, energy-dispersive X-ray spectroscopy indicates a decrease of the Sr content after acidic treatment. Using a 30 keV beam ($\approx\,$ 1$\mu$m penetration depth in Sr$_2$IrO$_4$), we find for both the virgin sample and the one soaked 3 hrs in 1 M KOH, Sr/Ir = 2.0 $\pm$ 0.02, and for the one soaked in 1 M H$_2$SO$_4$, Sr/Ir = 1.90 $\pm$ 0.02. Finally, there was no significant change of the cyclic voltamogramms in alkaline conditions, either for the sintered samples or the single crystal ones (supplement S4). Our observations confirm several studies pointing out that rare earth elements leach from the iridate perovskite in acidic media\cite{Diaz16,Seitz16}, this instability being closely related to the oxygen evolution activity of the iridate perovskites\cite{Grimaud16}.

As stated above, the saturation charge involved in the electrochemical oxidation of the single crystals exceeds the capacity of the available sites of a perfect surface, as in reaction \ref{evolution2}. As the computed oxygen variations are minute, the effect might concern only defects of the crystals. Although it is unlikely that such defects could induce bulk doping effects, as observed below for a thin film, some evidence for bulk structural modifications is valuable. While an absolute determination of the samples cristallographic parameters to better than a few 10$^{-3}$ \AA  was not possible from Rietveld refinements, due to a systematic skewness of the diffraction peaks for all powder samples, we could observe a systematic shift of the peaks of $\theta-2\theta$ scans before and after electrochemical oxidation, allowing for an estimate of the relative cristallographic parameters variation (supplement S5). Doing so, we find $dc/c$ = +8$\,\pm\, 1$ 10$^{-2}$/$\delta$ and $da/a$ = +1$\,\pm\, 1$ 10$^{-2}$/$\delta$. The \textit{c}-axis parameter variation is comparable to the ones in Refs.~\onlinecite{Grenier92,Radaelli93,Jorgensen89} reporting $dc/c$ = 4-7 10$^{-2}$/$\delta$, and $da/a$ = -4 10$^{-3}$ to -9 10$^{-2}$/$\delta$, for the cuprate and nickelate parent compounds, but well below the outstanding large value in Ref.~\onlinecite{Korneta10}, $dc/c$ = 0.3/$\delta$ for Sr$_2$IrO$_{4-\delta}$.

\begin{figure}
\resizebox{0.95\columnwidth}{!}{%
  \includegraphics{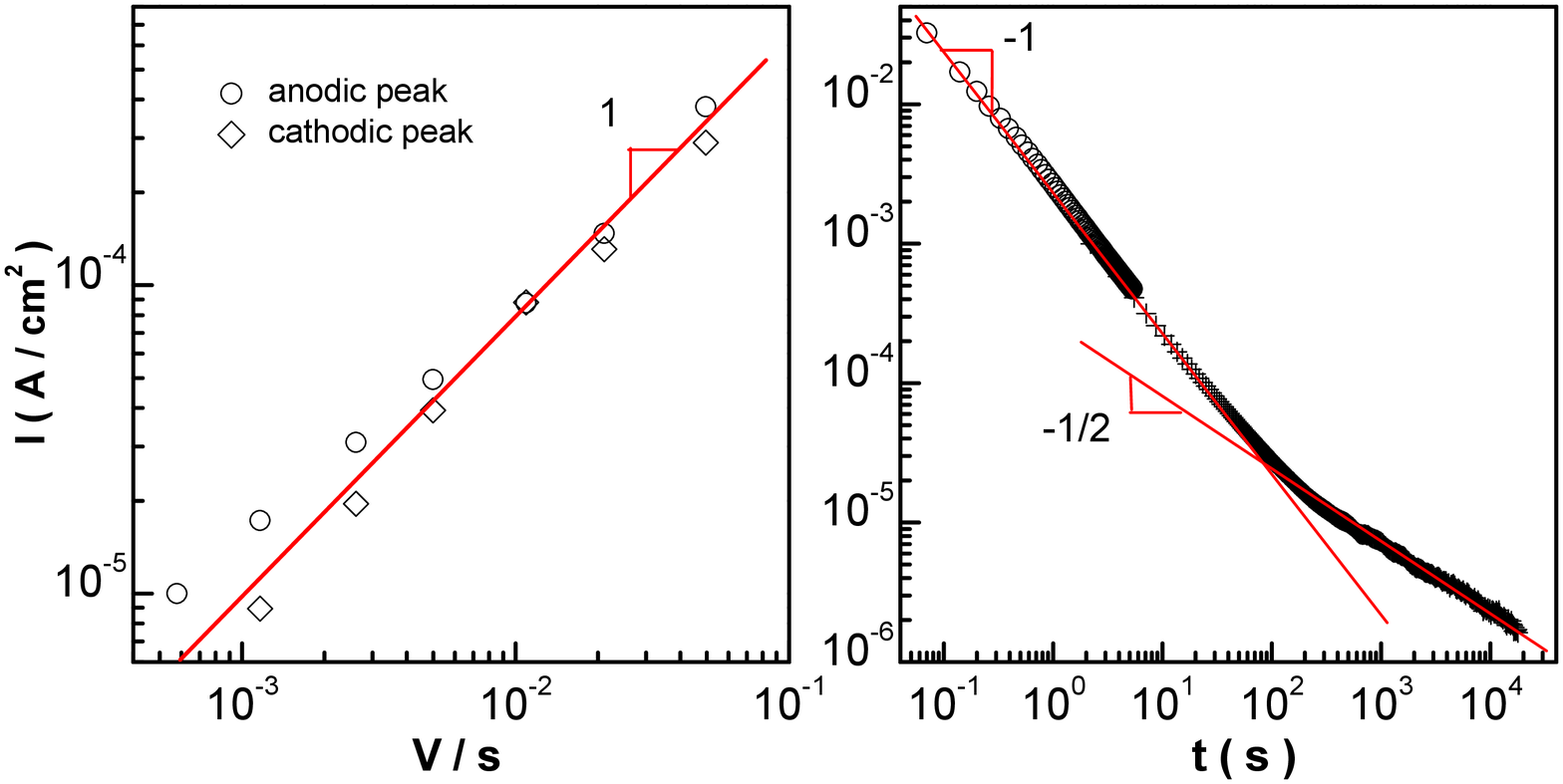}
  }
\caption{Left: anodic and cathodic peak current dependence on sweep speed (cleaved (001) Sr$_2$IrO$_4$ single crystal surface, in 1 M KOH). Right: relaxation of the cathodic current, after application of a step voltage from 460 mV to 110 mV (crosses and circles are obtained from two distinct acquisitions).}\label{cinetique}
\end{figure}

The diffusion length obtained above is much larger than can be anticipated from conventional diffusion. Indeed, a typical diffusion constant is obtained for La$_2$CuO$_4$, $D$ = $D_0 \exp{(-E_0/T)}$, where $D_0 \simeq$ 10$^{-6}$ cm$^2$/s and $E_0 \simeq$ 0.8 eV, yielding the ambient temperature diffusion coefficient $D \simeq$ 10$^{-20}$ cm$^2$/s\cite{Routbort94}. Thus, the conventional diffusion length, $\sqrt{Dt}$, cannot likely be reached within relaxation times of the order of 10$^3$ s, as observed here, and a faster mechanism must be considered. The investigation of the kinetics of the redox reactions provides insight into the actual diffusion mechanism. Fig.~\ref{cinetique} displays the anodic and cathodic peaks dependence on voltage sweep rate, and the relaxation of the cathodic current after a step voltage change. Clearly, the proportionality of the redox peaks with sweep rate (in place of the conventional $v^{1/2}$ dependence, for diffusion-limited mass transfer) is reminiscent of the absence of a mass transfer limitation, as is the case when only adsorbed species in quasi-equilibrium with the electrolyte are electroactive (for a review, see Ref.~\onlinecite{Bard}). Thus, the $v$ proportionality in Fig.~\ref{cinetique} originates from the nernstian equilibrium of an adsorbed layer, and does not require any kinetics contribution to the galvanic current. The latter are however perceptible as one focuses on the relaxation of the current (Fig.~\ref{cinetique}). We find two regimes. Besides a $t^{1/2}$ regime at long time (alternatively at low current) indicative of a limitation by diffusion, a short time $t^{-1}$ regime is observed.
 
In a general way, the kinetics for mass transfer is determined by \textit{i:} charging of the double layer, \textit{ii:} diffusion in the liquid, \textit{iii:} mass transfer in the solid electrode, which may involve both linear and non-linear diffusion. The first two mechanisms are too fast to limit mass transfer in the present case. Indeed, mechanism \textit{i:} may be described with an equivalent $RC$ circuit, with typical values $C$ $\simeq$ 10-100 $\mu F/cm^2$ and $R^{-1}$ $\simeq$ 0.1-10 $\Omega^{-1}/cm^2$ for a flat electrode, yielding typical time in the msec range, which is much larger than the time window (200 sec) observed for the $t^{-1}$ regime. We have checked that the high frequency impedance of a cleaved crystal electrode is indeed well described by the Cole equation, with characteristic time $\approx$ 1 msec (supplement S6). The Cottrell equation for mechanism \textit{ii:} gives for the current $i/A \simeq F D^{1/2} C^* / t^{1/2}$. Using $D$ = 10$^{-5}$ cm$^2$/s and $C^*$ = 1 M, we have, at $t$ = 10$^4$ s, $i/A$ = 3 10$^{-3}$ A/cm$^2$, which is two orders of magnitude larger than observed in the $t^{-1/2}$ regime. Thus, mass transfer is actually governed by the kinetics in the solid electrode, and the observed $t^{-1}$ and $t^{-1/2}$ regimes do not originate in the liquid double layer and diffusion mechanisms, respectively. We argue that the initial $t^{-1}$ regime may be explained by a strongly non-linear diffusion mechanism, and the $t^{-1/2}$ regime by classical diffusion in the solid. Indeed, it is reasonable to assume that strong concentration gradients, which are present at the electrode surface when diffusion in the electrode is negligible, strongly reduce the activation barrier for diffusion. This may be modeled by a diffusion constant, $D(C(x))$, which depends on a power law of the local gradient concentration in the solid. In this case, the diffusion equation in 1D for the concentration (we assume a semi-infinite electrode) becomes:
\begin{equation}
\frac{\partial C}{\partial t} \propto \left(\frac{\partial C}{\partial x}\right)^\alpha \frac{\partial^2 C}{\partial x^2}
\label{diffusion}
\end{equation}
where $\alpha$ is some exponent ($\alpha$ = 0, in the conventional situation). The physical origin of the power law behavior may be found in stress-assisted diffusion, where the driving force for diffusion is assumed a power law of the distance from interstitial defects\cite{Shewmon,Chou96}. It is easy to show that the conventional asymptotic regime for the mass flow rate, $J \propto t^{-1/2}$, extends in this case to $J \propto t^{-(1+\alpha)/(2+\alpha)}$. We believe the power law behavior for $D(C(x))$ is sufficiently general to allow for the validity in the $\alpha \gg 1$ regime, where $t^{-1}$ behavior is expected, independent of the $\alpha$ value. In this strongly non-linear regime, a quasi-linear concentration profile is expected at the electrode surface. The interpretation of the relaxation curve is then straightforward: at short times and strong concentration gradient, the non-linear regime prevails, associated to the $t^{-1}$ regime and to a strong reduction of the effective diffusion coefficient, followed by the conventional regime, associated to the $t^{-1/2}$ behavior. Thus, diffusion mostly occurs in the fast initial regime that allows a large diffusion length over short times, as estimated here.

\begin{figure}
\resizebox{0.95\columnwidth}{!}{%
  \includegraphics{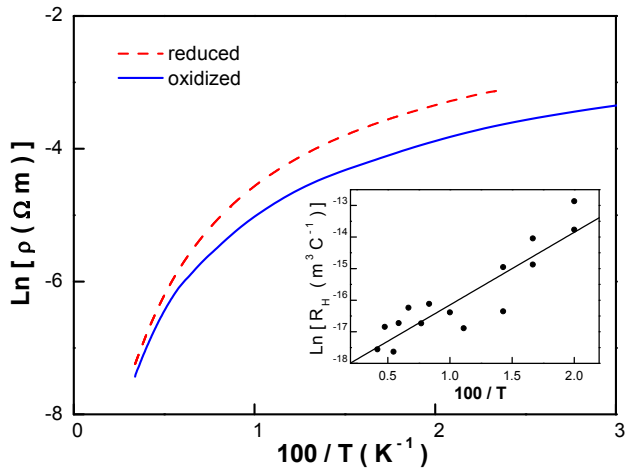}
  }
\caption{Resistivity of a 88 nm-thick film, after reduction (110 mV) and oxidation (460 mV) in 1 M KOH. The inset shows the Hall number, for the reduced film. The line is a fit to an Arrhenius law, with activation energy 20 meV.}\label{film}
\end{figure}

The influence of oxygen doping on transport properties could be evaluated, using the thin film sample. Fig.~\ref{film} shows that a reduction of the resistivity is obtained after introduction of the interstitial oxygen. From the charge involved during the cycling of the thin film, we obtain $\delta$ = 0.016. This is in agreement with the previous estimate from sintered samples, with a uniform oxygen doping concentration, as the film thickness is much smaller than the doping depth computed above. The carrier density could be evaluated from Hall measurements (Fig.~\ref{film}, inset). Despite a large noise (likely associated to the large resistance value of the sample), the data may be roughly fitted to an exponential behavior, with activation energy 20 meV. At ambient temperature, the Hall number for the reduced sample indicates hole doping with carrier concentration $p$ $\simeq$ 3 $10^{20}$ cm$^{-3}$, i.e. $\delta_p$ = 0.03 hole per formula unit, whereas at 50 K it is only 6 10$^{-4}$ (this is in line with the data in Ref.~\onlinecite{Klein2009}, for as-grown Sr$_2$IrO$_4$). Data in Fig.~\ref{film} show that resistivity curves are not simply shifted from one another, showing that the impurity levels associated to the intercalated oxygen are distinct from the ones that provide the carriers in the reduced material (we note that reduction coulometry for as-grown sintered samples indicates only a minute oxygen variation, $\delta$ = -5 10$^{-4}$, indicating a very small concentration of excess oxygen). At 50 K, if the interstitial oxygen fully transfers its charge to the valence band, and disorder does not vary too much, we would expect a decrease of the resistivity by a factor 2 $\delta / \delta_p \simeq$ 50. The observed variation is more than one order of magnitude smaller than this estimate. This weak charge transfer - as compared to the reduced sample -  suggests that the acceptor sites associated to the intercalated oxygen are not saturated at 50 K, as the impurity level resides far above the valence band (the ionization energy would be larger than 20 meV).

\begin{figure}
\resizebox{0.95\columnwidth}{!}{%
  \includegraphics{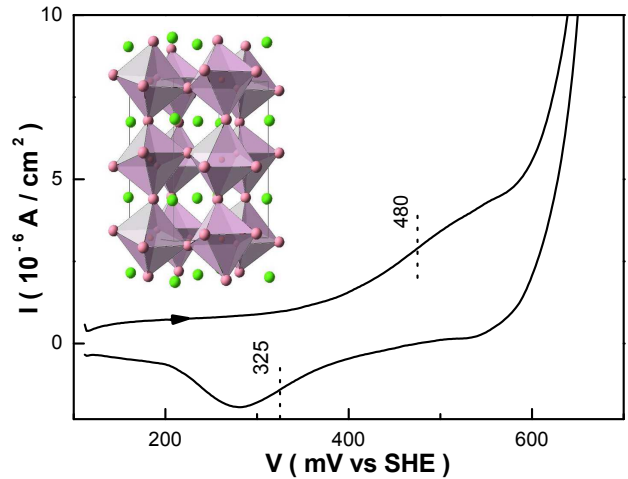}
  }
\caption{Cyclic voltammograms for a 64 nm-thick SrIrO$_3$ thin film ($S$ = 5.2 mm$^2$), in 1 M KOH (2.3 10$^{-3}$ V/s). The structure is from Ref.~\onlinecite{Zhao2008}.}\label{SrIrO3}
\end{figure}

The case of SrIrO$_3$ provides an interesting comparison with the layered compound Sr$_2$IrO$_4$. The former, being the 'infinite phase' of the Ruddlesden-Popper series, Sr$_{n+1}$Ir$_n$O$_{3n+1}$ (n = $\infty$), is essentially three-dimensional and does not contain any double SrO planes (Figs~\ref{structure},~\ref{SrIrO3}). Thus, there is no site which can reasonably accommodate the large O$^{2-}$ ion. Cyclic voltammograms on epitaxial thin films\cite{Li2016} are found essentially similar to the ones on Sr$_2$IrO$_4$ (Fig.~\ref{SrIrO3}). However, the charge involved in the redox process of the Ir$^{5+/4+}$ couple is found only 12 C cm$^{-3}$: this would correspond to the oxygen stoichiometry variation as small as $\delta$ = 2 10$^{-3}$. Rather than to a bulk reaction, the charge is here more likely associated to a surface one, as discussed in Ref.~\onlinecite{Tang16}. Indeed, the charge is found to correspond to 0.8 $e^-$ per Ir atom of a (110) surface, which is close to one elementary charge per Ir$_{surf}$, as in reaction \ref{evolution2}. This extreme case illustrates further the importance of steric effects for the possibility to electrochemically insert oxygen into these structures.

\begin{figure}
\resizebox{0.95\columnwidth}{!}{%
  \includegraphics{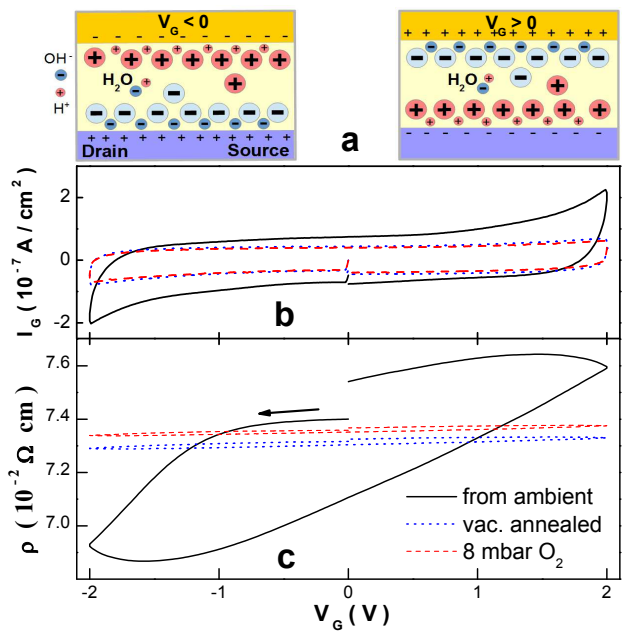}
  }
\caption{Gating with ionic liquid the 88 nm thick Sr$_2$IrO$_4$ film (b: gate current, c: film resistivity. $T$ = 220 K). Full line is for ionic liquid applied in ambient conditions, dotted line is for pumping first on ionic liquid at 300 K, 10$^{-5}$ mbar for one hour, both then measured in 10 mbar He thermal exchange gas. Dashed line is for pumped ionic liquid cooled down to 220 K in 8 mbar O$_2$. a: sketch of water dissociation upon gating (after Ref.~\onlinecite{Yuan10}). $dV_G/dt$ = 2 mV/s}\label{ionic}
\end{figure}

Finally, we show how the electrochemical intercalation of oxygen might also play a role in gating experiments of iridate perovskites which use ionic liquids (IL). Gating using highly polarizable ILs has been widely used to dope various materials, with induced surface charge typically one order of magnitude larger than can be obtained from gating through conventional insulators, as a result of the higher electric field obtained at the Helmholtz layer\cite{Misra07}. However, reversible oxygen migration from the gated material could be observed, accounting for the doping effect, in VO$_2$\cite{Jeong13}. In this case, the effect was attributed to the electric field-induced migration of the oxygen present in the IL. In Ref.~\onlinecite{Lu15}, an unexpected bulk gating effect for Sr$_2$IrO$_4$ films as thick 40 nm was proposed, whereas conventional field effect is expected to be screened over a few nm, in the case of oxides with carrier density a few 10$^{20}$ cm$^{-3}$\cite{Mannhart96}. The field-induced carrier density exceeded by a factor 30 the one expected from the IL capacitance.

We also observed a significant change of the resistivity of a thick film upon gating with DEME-TFSI IL\cite{DEME} (as measured in a Quantum Design PPMS cryostat, in 10 mbar He thermal exchange gas, using the geometry in Ref.~\onlinecite{Lu15}). The amplitude of the effect is smaller that in Ref.~\onlinecite{Lu15}, which reports a resistivity change by a factor 10 and 2, for 20 and 40 nm thick samples respectively, but we investigated here a thicker film (88 nm). Different surface barriers may also play a role in the amplitude of the effect. The sign of the effect is compatible with further hole-doping of the thin film upon negative gate biasing, as could be expected from conventional field effect (Fig.~\ref{ionic}c). Nevertheless, we find that this effect vanishes when the IL is pumped prior to measurement at 220 K. As the introduction of dry O$_2$ after pumping on the liquid does not restore the effect (Fig.~\ref{ionic}c), it is more likely due to dissolved water than dissolved oxygen, in the present case. Indeed, it is known that the introduction of an amphoteric molecule like water into the IL is able to produce surface hydrogenation or hydroxylation, depending on biasing\cite{Yuan10} (Fig.~\ref{ionic}a). The I$_G$-V$_G$ characteristic of the cell formed by the gate, the IL, and the Sr$_2$IrO$_4$ film is nearly ideally capacitive for the pumped IL, while it clearly develops a resistive component when not evacuated in vacuum (Fig.~\ref{ionic}b). This resistive component is the signature of Faradaic reactions at the electrodes in the presence of water, and it is probable that the doping of the film occurs by means of oxygen intercalation, as above. In view of the preceding considerations on material stability, it is expected that positive biasing, corresponding to acidic conditions at the film surface, can induce Sr extrusion and degradation of the film. Although such an electrochemical effect may not account for all the observations of iridate perovskites IL-doping as in Refs.~\onlinecite{Ravichandran13,Lu15}, it should be considered as a possible contribution in IL-doping investigations.

To summarize, we have found that oxygen electrochemical intercalation of oxygen into Sr$_2$IrO$_4$, although more efficient than obtained from conventional gas process, only yields small oxygen stoichiometry variation, $\delta$ $\simeq$ 0.01. The diffusion of oxygen is likely assisted by the large concentration gradient upon electrochemical cycling, and extends over a few $\mu$m from single crystal surface. This small oxygen concentration may be due to the large steric effects associated to Sr$^{2+}$. We find that the hole doping contribution from the extra oxygen to the resistivity is weak, indicating a poor charge transfer to the conduction band. The possibility for similar electrochemical effects in gating experiments, using water-contaminated ionic liquids, was shown.

\section*{}
We acknowledge support from the Agence Nationale de la Recherche grant SOCRATE ANR-15-CE30-0009-01.


\end{document}